\documentclass{llncs}

\usepackage{tikz}
\usetikzlibrary{matrix,chains,positioning,decorations.pathreplacing,arrows}

\usepackage{amsmath}
\usepackage{subcaption}

\usepackage{array}
\newcolumntype{M}[1]{>{\centering\arraybackslash}m{#1}}

\usepackage{booktabs}
\usepackage{amsfonts}
\usepackage{siunitx}
\usepackage{soul}
\usepackage{wrapfig}
\usepackage{enumitem}
\usepackage{capt-of}

\begin{document}

\title{Social Influence (Deep) Learning \\ for Human Behavior Prediction}
\author{Luca Luceri \inst{1}\inst{,}\inst{2}, Torsten Braun \inst{2}, Silvia Giordano \inst{1}}
\institute{University of Applied Sciences and Arts of Southern Switzerland (SUPSI)
\and University of Bern \\
\email{luca.luceri@supsi.ch, braun@inf.unibe.ch, silvia.giordano@supsi.ch}}

\maketitle

\begin{abstract}
\vspace{-.5cm}
Influence propagation in social networks has recently received large interest.
In fact, the understanding of how influence propagates among subjects in a social network opens the way to a growing number of applications. Many efforts have been made to quantitatively measure the influence probability between pairs of subjects. 
Existing approaches have two main drawbacks: $(i)$ they assume that the influence probabilities are independent of each other, and $(ii)$ they do not consider the actions not performed by the subject (but performed by her/his friends) to learn these probabilities.
In this paper, we propose to address these limitations by employing a deep learning approach.
We introduce a Deep Neural Network (DNN) framework that has the capability for both modeling social influence and for predicting human behavior.
To empirically validate the proposed framework, we conduct experiments on a real-life (offline) dataset of an Event-Based Social Network (EBSN).
Results indicate that our approach outperforms existing solutions, by efficiently resolving the limitations previously described.
\vspace{-.3cm}
\end{abstract}

\section{Introduction}
Influence propagation in social networks has recently received large interest, both in academia and industry.
In fact, the understanding of how influence propagates in a social network opens the door to a wide range of applications, as targeted advertising, viral marketing, and recommendation.
In this context, social networks play an important role as a medium for spreading processes \cite{newman2003structure,albert2002statistical}. 
As an example, a new idea can spread through a social network in the form of ``word-of-mouth'' communication \cite{goldenberg2001talk}. 
In the last decade, particular attention has been devoted to the comprehension and modeling of the social influence phenomenon. 
Social influence is recognized as a key factor that governs human behavior.
It indicates the attitude of certain individuals to be affected by other subjects' actions and decisions.
The idea is that the interaction with other individuals (or a group) may result in a change of subject's thoughts, feelings, or behavior. 
In other words, a subject may take a decision, e.g., to buy a new product or to watch a TV show, when she/he sees her/his friends taking that decision. 

A considerable amount of work has been conducted to investigate social influence and analyze its effect.
In \cite{singla2008yes} and \cite{anagnostopoulos2008influence}, the authors propose how to qualitatively measure the existence of social influence, whereas in \cite{crandall2008feedback} the correlation between social similarity and influence is examined.
In \cite{luceri2017communities}, we introduce a novel interpretation of physical, homophily, and social community, as sources of social influence.
Other relevant works focused on the problem of influence maximization \cite{domingos2001mining,richardson2002mining,kempe2003maximizing,kimura2007extracting}.
This problem aims to find the most influential individuals in a social network in order to maximize the number of influenced subjects.
Viral marketing is a strategy that exploits this idea to promote new products. 
Kempe et al. \cite{kempe2003maximizing} focus on two fundamental propagation models, referred to as Independent Cascade (IC) model and Linear Threshold (LT) model.
In the IC model, each subject independently influences her/his friends with given influence probabilities. 
In the LT model, a subject is influenced by her/his friends if the combination of their total influence probabilities exceeds a threshold. 
Both models assume to have as input a social network whose edges are weighted by a measure of influence probability. However, these values are not known in practice and, thus, should be estimated. 
Many efforts have been made to quantitatively measure the influence strength between pairs of friends \cite{gruhl2004information,saito2008prediction,tang2009social,goyal2010learning,liu2012learning,fang2013predicting}. 
In particular, Goyal et al. \cite{goyal2010learning} and Saito et al. \cite{saito2008prediction} investigate how to learn the influence probabilities using only the history of subjects' actions.
Such approaches have two main drawbacks: $(i)$ they assume that the probability of friends influencing a subject are independent of each other, and $(ii)$ they do not consider the actions not performed by the subject (but performed by her/his friends) to learn the influence probabilities. 

In this paper, we propose to address the aforementioned drawbacks by employing a deep learning approach.
Our objective is to learn subjects interplay for modeling social influence and predicting their behavior.
We summarize our contributions as follows:
\begin{itemize}
\item We introduce a Deep Neural Network (DNN) framework that has the capability for both learning social influence and predicting human behavior. 
To the best of our knowledge, our solution is the first architecture that accomplishes these two tasks in one shot. 
\item We model social influence among subjects overcoming the assumptions introduced by previous works.
We design a DNN taking into account both $(i)$ the relationship between the subject and her/his friends and $(ii)$ the interactions among them. 
Further, we learn social influence considering also the actions not performed by the subject (but performed by her/his friends) to understand who really affects subject's decisions.
\item We evaluate the performance of our approach using data from an Event-Based Social Network (EBSN). 
This allows us to investigate social influence considering together \emph{online} (through the social network) and \emph{offline} (real-life) social interactions.
Previous works conducted their experiments analyzing social influence only in Online Social Networks (OSNs).
We compare our approach with existing solutions, achieving a remarkable improvement. 
\end{itemize}


\section{Problem Definition}
\label{pro_def}
In this paper, we aim to learn social influence in a social network in order to predict human behavior, in terms of decision and actions performed by individuals.
Let $G=(V,E)$ be a directed graph, which represents the social network, where $V=\{u_1,u_2, \dots, u_N\}$ is the set of subjects and $E$ is the set of edges connecting them. 
Subject $u_j$ is considered a \emph{friend} of subject $u_i$ if $(u_j,u_i) \in E$.
To model social influence we measure the strength of friends' influence on subject's actions.
We define $A$ as the whole set of actions. 
For each action $a \in A$, each subject is either \emph{active}, if she/he has performed the action, or \emph{inactive}, otherwise. It should be noticed that inactive subjects may become active, but not the opposite.
We define $S_{u_i,a}$ as the set of active friends of $u_i$ for the action $a$.
The objective is to predict whether a subject becomes active based on her/his active friends.
To achieve this purpose, previous works determine the influence probability $p_{u_i}(S_{u_i,a})$, i.e., the influence exerted on subject $u_i$ by the active friends $S_{u_i,a}$, by exploiting the history of $u_i$ actions.
The main assumption in these works is that the probability of various friends influencing $u_i$ are independent of each other.
Thereby, the probability $p_{u_i}(S_{u_i,a})$ is computed as
$p_{u_i}(S_{u_i,a})=1-\prod_{u_j\in S_{u_i,a}}(1-p_{u_j,u_i})$,
where $p_{u_j,u_i}$ is the influence probability of $u_j$ on $u_i$.

\begin{figure*}
	\vspace{-.3cm}
        \centering
        \resizebox{0.6\textwidth}{!}{%
        \begin{subfigure}[b]{0.47\textwidth}
        \centering
               \begin{tikzpicture}[auto, node distance=2cm, every loop/.style={},
                    thick,node/.style={circle,draw},other node/.style={circle,draw,red,dashed}, main node/.style={circle,draw, dashed}]

  \node[main node] (1) {\Large $u_1$};
  \node[node] (5) [below of=1] {\Large $u_5$};
  \node[main node] (2) [left of=5] {\Large $u_2$};
  \node[other node] (3) [below of=5] {\Large $u_3$};
  \node[main node] (4) [right of=5] {\Large $u_4$};

  \path[every node/.style={font=\sffamily\large}, ->]
    (1) edge node [right] {0.7} (5)
    (2) edge node {0.4} (5)
    (3) edge node [right] {0.8} (5)
    (4) edge node {0.2} (5);
\end{tikzpicture}

                \caption{}
                \label{ego}
        \end{subfigure}\hfill%
        \begin{subfigure}[b]{0.47\textwidth}
        \centering
              \begin{tikzpicture}[auto, node distance=2cm, every loop/.style={},
                   thick,node/.style={circle,draw},other node/.style={circle,draw,red,dashed}, main node/.style={circle,draw, dashed}]

  \node[main node] (1) {\Large$u_1$};
  \node[node] (5) [below of=1] {\Large$u_5$};
  \node[main node] (2) [left of=5] {\Large$u_2$};
  \node[other node] (3) [below of=5] {\Large$u_3$};
  \node[main node] (4) [right of=5] {\Large$u_4$};

  \path[every node/.style={font=\sffamily\large}, ->]
    (1) edge node [right] {0.7} (5)
	edge node[right] {0.8} (2)
    (2) edge node {0.4} (5)
    	 edge [bend left]  node[left] {0.7} (1)
    (3) edge node [right] {0.8} (5)
    (4) edge node {0.2} (5);
\end{tikzpicture}

                \caption{}
                \label{ego2}
        \end{subfigure}\hfill  
}%
        \caption{Example of influence probabilities in a social network}
	\label{ego_all}
	\vspace{-.5cm}
\end{figure*}
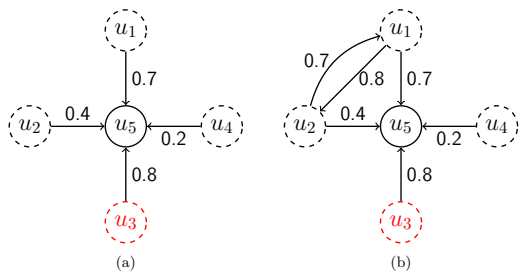

As an example, Figure \ref{ego} represents the social network of subject $u_5$.
To simplify the reading, only the incoming edges of node $u_5$ are represented.
Each edge is weighted by the influence probability $p_{u_j,u_i}$. A red node represents an inactive subject.
The decision of $u_5$ to perform an action $a$ is a function (Eq. (1)) of the active friends ($u_1,u_2,u_4$) and related influence probabilities. 

Existing approaches learn the probability $p_{u_j,u_i}$, $\forall (u_j,u_i) \in E$, from the actions performed by both $u_j$ and $u_i$. 
In particular, they consider $u_i$ as influenced by $u_j$ if the latter performed the action before the former.
Such approaches have two main drawbacks. 
The probability of friends influencing a subject are considered independent of each other. This assumption may not be always true, especially when two friends of a subject are in turn friends, as for the nodes $u_1$ and $u_2$ in the example of Figure \ref{ego2}. 
The fact that subject $u_1$ and $u_2$ are both active can differently affect subject $u_5$ decision.
In this instance, the joint probability of influencing $u_5$ should be higher if compared to the combination of the independent probabilities (Eq. (1)). 
Further, previous works in the literature learn the influence probability by considering only the actions performed by the subject (\emph{positive samples}).
However, it may be relevant to take into account the actions not performed by the subject (\emph{negative samples}), but performed by her/his friends, so as to understand who really affects subject's decisions. 
As an example, we consider the scenario where subject $u_5$ does not buy a certain product, while some of her/his friends do. In this instance, considering also negative samples can improve the influence modeling, as $u_5$ may be affected by the friends that share the same \emph{negative} decision.

Previous works differ from each other for the way the probabilities $p_{u_j,u_i}$ are estimated.
In this paper, we study the LT models proposed by Goyal et al. \cite{goyal2010learning} and the IC model of Saito et al. \cite{saito2008prediction}. 
Other works in the literature model social influence at topic-level, i.e., considering influence among subjects with respect to a set of OSN topics.
We are not only interested in online scenarios, thus, we aim to model social influence among subjects independently of the topics.
In the LT models of Goyal et al., a node becomes active if $p_{u_i}(S_{u_i,a}) \geq \theta$, where $\theta$ is the activation threshold. 
They propose different probabilistic models to capture the influence probability $p_{u_j,u_i}$, referred to as Bernoulli Distribution (BD), Jaccard Index (JI), Partial Credits - Bernoulli (PC-B), and Partial Credits - Jaccard (PC-J). We do not describe them in details for the lack of space.
In the IC model of Saito et al., each active subject independently influences her/his inactive friends with influence probabilities estimated by maximizing a likelihood function with the Expectation Maximization (EM) algorithm.

\section{Proposed Solution}
This work addresses the aforementioned drawbacks by formalizing a deep learning approach for modeling social influence and predicting subject's behavior.
In this section, we present the proposed approach based on a DNN architecture.


\subsection{Deep Neural Network (DNN)}
In recent years, deep learning \cite{lecun2015deep,schmidhuber2015deep} has found successful application in a growing number of areas. 
A DNN is able to approximate any continuous function by learning the relationships embedded in the input data. 
Thereby, it replaces the manual feature extraction procedure by building up a complex hierarchy of concepts through the multiple layers of the network to automatically extract discriminative and abstractive features of data \cite{he2017neural}.
A DNN is defined by a combination of three layers: input layer $(\mathbf{x})$, hidden layers $(\mathbf{h}_1,\mathbf{h}_2, \dots, \mathbf{h}_L)$, and output layer $(\mathbf{y})$.
These layers are fully connected in a weighted way as follows
\vspace{-0.1cm}
\[
    \mathbf{h}_j= 
\begin{cases}
    \phi_j(\mathbf{x}\mathbf{W}_{xh_{j}})& \text{if } j= 1 \\
    \phi_j(\mathbf{h}_{j-1}\mathbf{W}_{h_{j-1}h_{j}})& \text{if } 1<j\leq L
\end{cases} 
\vspace{-0.2cm}
\]
\begin{equation*}
\mathbf{y}= \phi_o(\mathbf{h}_{L}\mathbf{W}_{h_{L}y}) \ ,
\end{equation*}
where $\mathbf{W}_{kl}$ indicates the weights of the connections between layer $k$ and $l$, while $\phi_j$ is a non-linear activation function (e.g., sigmoid, ReLU, tanh, softmax) of each hidden node at layer $j$, and $\phi_o$ is a non-linear activation function of each output node.
The predictive model of a DNN can be formulated as $\hat{\mathbf{y}}=f(\mathbf{x}|\Theta)$, where 
$\hat{\mathbf{y}}$ denotes the predicted output, $\Theta$ represents the model parameters (i.e., the inter-layers weights), and $f$ indicates the function that maps the input $\mathbf{x}$ to the output $\hat{\mathbf{y}}$ based on the DNN architecture, i.e., $f(\mathbf{x})=\phi_o( \phi_{L}(\dots\phi_{2}(\phi_{1}(\mathbf{x}))\dots))$.

\subsection{Social Influence Deep Learning}
\begin{wrapfigure}{r}{0.4\textwidth}
 \vspace{-.8cm}
    \resizebox{0.4\textwidth}{!}{%
\begin{tikzpicture}
[   cnode/.style={draw=black,fill=#1,minimum width=3mm,circle},
]

    \node[cnode=red,label=0:\Large $\hat{y}_{u_i,a}$] (s) at (5,-5.4) {};
    \node at (1,-4) {$\vdots$};
     \node at (1,-9) {$\vdots$};
    \node at (3,-5.5) {$\vdots$};
    
    \foreach \x in {1,...,4}
    {   \pgfmathparse{\x<4 ? \x : "N"}
        \node[cnode=blue,label=180:\Large $x_{\pgfmathresult}$] (x-\x) at (1,{-\x-div(\x,4)}) {};
    }
    
        \node[cnode=blue,label=180:\Large $x_{N+1}$] (x-5) at (1,{-5-1-div(1,4))}) {};
        \node[cnode=blue,label=180:\Large $x_{N+2}$] (x-6) at (1,{-5-2-div(2,4))}) {};
        \node[cnode=blue,label=180:\Large $x_{N+3}$] (x-7) at (1,{-5-3-div(3,4))}) {};
      	\node[cnode=blue,label=180:\Large $x_{2N}$] (x-8) at (1,{-5-2-div(12,4))}) {};
      \draw [decorate,decoration={brace,amplitude=10pt},xshift=4pt,yshift=-3.5cm] (-0.8,-1.9) -- (-0.8,2.8) node [black,midway,xshift=-1.4cm] {\Large $\mathbf{v}_{u_i}^U$};
      \draw [decorate,decoration={brace,amplitude=10pt},xshift=4pt,yshift=-5.5cm] (-0.8,-4.8) -- (-0.8,-0.2) node [black,midway,xshift=-1.4cm] {\Large $\mathbf{v}_{u_i}^{F_{a}}$};

    \foreach \x in {1,...,6}
    {   \pgfmathparse{\x<6 ? \x : "m"}
        \node[cnode=gray,label=0:\Large $h_{\pgfmathresult}$] (p-\x) at (3,{-1.5-\x-div(\x,4)}) {};
        \draw (p-\x) -- node[above,sloped,pos=0.3] {} (s);
    }

    \foreach \x in {1,...,8}
    {   \foreach \y in {1,...,6}
        {   \draw (x-\x) -- (p-\y);
        }
    }     
 \end{tikzpicture}
 }%
  \vspace{-.1cm}
 \caption{DNN Framework}
 \label{dnn}
 \vspace{-.6cm}
\end{wrapfigure}
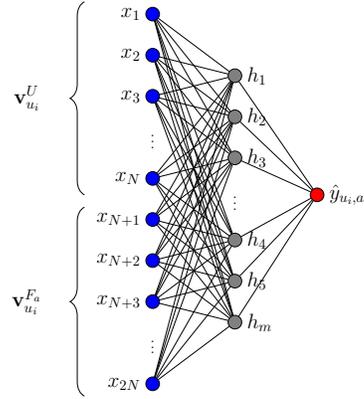
In this work, we address the limitations of existing approaches by learning the interplay among subjects using a DNN.
The proposed approach has the capability for both modeling social influence and predicting human behavior in one shot.
It should be noticed that the DNN does not explicitly produce a mathematical model, but it learns abstractive feature to implicitly model and learn the interaction of the data in input.
Our task can be formulated as the problem of predicting whether subject $u_i$ performed action $a$ as a function of the active friends $S_{u_i,a}$.
We address this task as a binary classification problem.
Thereby, the output $y_{u_i,a}$ of the DNN is a Boolean variable that is equal to 1 if $u_i$ performed $a$, and is 0 otherwise.
The input layer consists of two vectors $\mathbf{v}_{u_i}^U$ and $\mathbf{v}_{u_i}^{F_{a}}$ that indicate subject $u_i$ and her/his active friends for the action $a$, respectively.
Both of them have length $N=|V|$.
The former is a one-hot vector that uniquely identifies each subject $u_i \in V$.
The vector consists of all \emph{zeros} with the exception of a single \emph{one} that identifies one element of the set.
In this instance, subject $u_i$ is represented by the vector $\mathbf{v}_{u_i}^U$, which has only the $i^{th}$ element equal to one.
The latter represents the active friends of subject $u_i$ for the action $a$.
The $j$-th element of $\mathbf{v}_{u_i}^{F_{a}}$ corresponds to subject $u_j$ and it equals one only if $u_j$ is active and $(u_j,u_i) \in E$, otherwise is equal to 0.
%
These two vectors are first concatenated and then fed into a multi-layer architecture, as depicted in Figure \ref{dnn}.
For the sake of simplicity, a DNN with only one hidden layer ($L=1$) is depicted. 
In our experiments, we design a network with a tower structure, where the bottom layer is the largest and the number of nodes of each successive layer is half of its precedent.
In such a way, higher layers with few nodes can learn more abstractive features from the input data \cite{he2016deep,he2017neural}.
Details about the implementation will be given in Section \ref{impl}.
The training is performed by minimizing the binary cross-entropy loss between $\hat{y}_{u_i,a}$ and $y_{u_i,a}$, where $\hat{y}_{u_i,a}=f(\mathbf{v}_{u_i}^U,\mathbf{v}_{u_i}^{F_{a}}|\Theta)$ is the predicted output of our DNN framework.

The rationale of this approach is based on the attempt of overcoming the drawbacks of previous works described in Section \ref{pro_def}.
We model social influence by considering the inter-dependencies among friends. In fact, according to the DNN architecture presented above, we take into account both $(i)$ the relationship between the subject and her/his friends and $(ii)$ the interactions among them.
We accomplish this task by placing the social network in a neural network, letting the DNN learn the influence strengths and the interplay among the subjects in the social network.
We learn social influence including in the training phase also actions not performed by the subject. 
For each subject, we train our DNN with an equal number of positive ($y_{u_i,a}=1$) and negative samples ($y_{u_i,a}=0$).
In such a way, the DNN framework has the capability for both modeling social influence and predicting human behavior in one shot. 

\section{Experimental Evaluation}
To empirically evaluate our framework, we conduct experiments using data of an EBSN.
This dataset allows us to investigate social influence considering both \emph{online} (through the social network) and \emph{offline} (real-life) social interactions.

\subsection{Dataset Description}
An EBSN is a web platform where users can create events, promote them, and invite friends to participate.
Events range from small get-together activities, e.g., Sunday brunch or movie night, to bigger events, e.g., concerts or conferences \cite{liu2012event}. 
The rationale behind the choice of utilizing an EBSN is based on the intrinsic agglomerative power of the events. 
In fact, participating in an event represent a direct and explicit form of social interaction, other than a personal interest.
An EBSN provides a social network service so as to connect friends and users with common interests.
In the event main page, a user can see the information related to the event, e.g., date, location, and description, along with the confirmed participants. This information may activate processes of social influence, which can drive user participation in the events \cite{georgiev2014call}.

In this study, we use a dataset from \emph{Plancast}, an EBSN for sharing upcoming plans with friends.
Plancast allows users to subscribe each other providing direct connections among them. 
Subscription is similar to the concept of \emph{following} in OSNs, e.g., Twitter. 
We utilize a dataset \cite{liu2012event} that includes 93041 users and 401634 events, combined in 1702058 user subscriptions and 869200 user-event participations.
We restrict our analysis to the U.S., as most of the events have been organized there. We filter out users without any subscription and that attended less than 20 events. 
We set this threshold in order to build, per each user, a reasonable training and test set to predict her/his behavior. 



\subsection{DNN Implementation}
\label{impl}
In this section, we describe how we implement and design our DNN framework.
The actions set  $A$ is defined by the user-events participation in the EBSN dataset, while $A_{u_i} \subseteq A$ is the set of events attended by subject $u_i \in V$. 
A subject is considered active for the event $a$ if she/he decided to participate in the event $a \in A$.
For each subject $u_i$, we randomly select $n_{u_i}$ events not attended by $u_i$ so as to consider also negative samples, where $n_{u_i}= |A_{u_i}|$.
In order to limit overfitting and to reduce variability, we utilize a 10-fold cross validation to split the dataset into training and test set. We build the folds so as to
preserve the percentage of positive and negative samples for each subject in the dataset.

We implement our DNN framework in Keras \cite{chollet2017keras}, following a tower pattern composed of $L=3$ layers with $\{128,64,32\}$ nodes, respectively. We train the network for 25 epochs using RMSProp as optimization function, employing the ReLu as activation function at the hidden layers and the sigmoid as activation function at the output layer.
Moreover, we apply a dropout technique, with a dropout equal to 0.1, to avoid overfitting.
We tune these hyper-parameter performing a grid search on a validation set (10\% of the data).

\subsection{Performance Comparison}
To validate the performance of our approach, we compare our proposed method (DNN) with the following baseline algorithms: the LT models (BD, JI, PC-B, and PC-J) proposed by Goyal et al. \cite{goyal2010learning}, and the IC model of Saito et al. \cite{saito2008prediction}.
To find the best threshold $\theta$ in the LT model, we measured two metrics: the Youden's index and the closest point to (0,1) in the Receiver Operating Characteristic (ROC) curve. We show only the performance related to the Youden's index as it achieves better results.
To examine the performance of these models, we employ widely used metrics in the evaluation of classification problem: Accuracy, True Positive Rate (TPR), and False Positive Rate (FPR).

\begin{table}[t]
\centering
\caption{Prediction performances comparison: DNN vs. LT models vs. IC model}
\label{res}       
%
%
\begin{tabular}{M{1.3cm}M{1.2cm}M{1.2cm}M{1.2cm}M{1.2cm}M{1.2cm}M{1.2cm}} 
\hline\noalign{\smallskip}
 & DNN & BD & JI & PC-B & PC-J & IC  \\
 \hline\noalign{\smallskip}
Accuracy & 85\% & 78\% & 77\%& 78\%& 77\% & 77\% \\    
TPR & 75\% & 74\% & 75\%& 66\%& 61\% & 60\% \\
FPR & 5\% & 14\% & 15\%& 6\%& 5\% & 5\% \\
\noalign{\smallskip}\hline\noalign{\smallskip}
\end{tabular}
\vspace{-.5cm}
\end{table}
Table \ref{res} depicts the performance of the different solutions.
Results indicate that the DNN framework achieves the best Accuracy, TPR, and FPR.
We empirically show that the proposed approach outperforms the baseline algorithms, by efficiently resolving the limitations related to the existing works.
This result highlights the importance of $(i)$ the interplay among subject's friends, in terms of dependent influence probabilities, and of $(ii)$ the negative samples to detect influential friends and learn influence strengths. 
Our DNN framework has the capability for both modeling social influence taking into account these aspects and for predicting human behavior, achieving remarkable results.
\section{Conclusions}
In this paper, we investigated social influence and how it impacts human behavior.
We propose to address the limitations of existing approaches by employing a deep learning approach.
We introduced a DNN framework that has the capability for both modeling social influence and predicting human behavior.
We implemented an architecture that allows the DNN to learn the interplay among friends and to consider both positive and negative samples.
To empirically validate the proposed framework, we evaluated our approach using real-life data of an EBSN.
Performances exhibit a significant improvement with respect to the state of the art, showing that the proposed approach efficiently resolves the limitations related to existing works.
\vspace{-.25cm}

\bibliographystyle{splncs.bst}
\bibliography{Paper203}

\end{document}